# Recalibration of the $H_{-0.5}$ Magnitudes of Spiral Galaxies

Giuseppe Tormen

Department of Astronomy, University of Padova, 35100 Padova, Italy

and

David Burstein

Department of Physics and Astronomy, Box 871504
Arizona State University, Tempe, AZ 85287-1504

## Abstract

The H magnitude aperture data published by the Aaronson et al. collaboration over a 10-year period is collected into a homogeneous data set of 1731 observations of 665 galaxies. 96% of these galaxies have isophotal diameters and axial ratios determined by the Third Reference Catalogue of Bright Galaxies (de Vaucouleurs et al. 1991), the most self-consistent set of optical data currently available. The precepts governing the optical data in the RC3 are systematically different from those of the Second Reference Catalogue (de Vaucouleurs, de Vaucouleurs and Corwin 1976), which were used by Aaronson et al. for their original analyses of galaxy peculiar motions. This in turn leads to systematic differences in growth curves and fiducial H magnitudes, prompting the present recalibration of the near-infrared Tully Fisher relationship. New optically-normalized H magnitude growth curves are defined for galaxies of types S0 to Im, from which new values of fiducial H magnitudes, $H^g_{-0.5}$, are measured for the 665 galaxies. A series of internal tests show that these four standard growth curves are defined to an accuracy of 0.05 mag over the interval $-1.5 \leq \log(A/D_g) \leq -0.2$. Comparisons with the Aaronson et al. values of diameters, axial ratios and fiducial H magnitudes show the expected differences, given the different definitions of these parameters. The values of $H^g_{-0.5}$ are assigned quality indices; a value of 1 indicates an accuracy of <0.2 magnitude; Quality 2 indicates an accuracy of 0.2 to 0.35 mag; and Quality 3 indicates >0.35 mag. Revised values of corrected HI velocity widths are also given, based on the new set of axial ratios defined by the RC3.

## 1. Introduction

After the initial discovery of the relationship between absolute B magnitudes and rotation velocity widths for spiral galaxies by Tully & Fisher (1977), Aaronson, Huchra and Mould (1979) showed that using near-infrared magnitudes, especially in the H (1.6$\mu$) passband, offered a tighter relationship. In the ensuing decade, this group, along with their colleagues used this "near-infrared Tully-Fisher" (IRTF) relationship to study the large-scale velocity field defined by the peculiar velocities of over 550 spiral galaxies (Aaronson 1977; Aaronson, Huchra and Mould 1979; Aaronson, Mould and Huchra 1980; Mould, Aaronson and Huchra 1980; Aaronson et al. 1980, 1981, 1982a,b, 1986, 1989; Aaronson and Mould 1983, 1986; Aaronson 1986; Cornell et al. 1987; Bothun et al. 1984, 1985; Bothun and Mould 1987). The corpus of work defined by these papers (designated in the reference list as A1 through A16) is referred to in the present paper as "Aaronson et al." The early papers in the Aaronson et al. series formed the first large dataset on the peculiar velocities of nearby galaxies. Their results remain as one of the cornerstones of our current understanding of the large-scale velocity field within a distance of 3000 km s$^{-1}$, as well as in specific regions of the sky at larger distances (Burstein 1990; these data also form the bulk of the data analyzed by Shaya, Tully and Pierce 1992).

Aaronson et al. based their fiducial H magnitude estimates for galaxies on H band aperture magnitudes combined with optical diameters and axial ratios. The hybrid magnitude that results from this combination of data, termed $H^0_{-0.5}$ by this group, is dependent on the cumulative accuracy of the H magnitudes, the optical isophotal diameters and the optical axial ratios (on which corrections for inclinations to the line-of-sight and internal extinction are based). In addition, statistical corrections must be done for Galactic extinction and redshift-dependent effects.

The decade in which the Aaronson et al. papers were published saw the supplanting of photographic magnitudes and photoelectric magnitudes with CCD-acquired data, as well as the beginnings of the now-burgeoning field of near-infrared imaging arrays. As noted by Aaronson et al. in their own papers (Aaronson et al. 1986; Aaronson et al. 1989), different observational methods of measuring optical diameters and inclinations of galaxies led them to continually try to assess the accuracy of their data, and may have led to differences in definitions of optical diameters among different sets of their data.

Separately, one of us (DB) has been engaged in a long-term effort to place the published peculiar motions of galaxies on an internally homogeneous system. This necessarily involves combining data sets from different groups that use different techniques to estimate distances of galaxies. In the process of assembly of the *Mark II Catalog of Galaxy Peculiar Velocities* (privately distributed by DB over e-mail since 1989), it was realized that the original distances of spiral galaxies published by Aaronson et al. (1982), mostly non-cluster galaxies, could be objectively divided into two subsets that differ significantly in accuracy (Faber & Burstein 1988; Burstein 1990; Burstein, Faber & Dressler 1990). The objective criterion used for this distinction was the accuracy of the optical data from which diameters and inclinations were derived (Faber & Burstein 1988). In addition, the Aaronson et al. cluster data (1986; 1989) were derived from a mixture of catalog and CCD data and, as such, it would be worthwhile to reanalyze these data with a homogeneous set of diameters.

Once it was decided to produce an updated (Mark III) catalog of galaxy peculiar velocities, it was deemed highly desirable to place all of the Aaronson et al. data set, field and cluster galaxies on a common optical system of diameters and axial ratios. Fortunately, at about the same time, the *Third Reference Catalog of Bright Galaxies* (de Vaucouleurs et al. 1991; hereafter RC3) was published, in which optical data for 96% of the galaxies studied by Aaronson et al. were placed on the same diameter/axial ratio system (the authors are grateful to H.G. Corwin, Jr. for providing the RC3 in computer-readable form).

Necessarily, a reanalysis of the Aaronson et al. data requires combining the H aperture measurements with the new optical data, rederiving standard growth curves, and producing new $H_{-0.5}$ magnitudes. One desirible offshoot of this recalibration was that it permits one to estimate the quality of each $H_{-0.5}$ magnitude for cluster and field galaxies based on a uniform set of objective criteria. These criteria include quality of optical data, quality of H magnitude data and the number of H aperture measurements with different aperture sizes (some galaxies have only one aperture measurement). Overall, our goal was to produce a data set for the H magnitudes of spirals the accuracy of which could be assessed in analogy with what was done for elliptical galaxy photometry by the 7 Samurai collaboration (Burstein et al. 1987; Faber et al. 1989).

The present paper is one of two papers that result from this recalibration effort. Here we assemble all of the available H magnitude aperture measurements for spiral galaxies and use these data to derive new $H_{-0.5}$ magnitudes for 665 field and cluster galaxies for which sufficient optical and H magnitude data exist. The intent of this work is to recast the Aaronson et al. survey data into a self-consistent set of measurements among the field data, and between field and cluster data. No attempt has been made to supplement either the actual H magnitudes or the HI velocity widths with more recent data. In the second paper (Willick, Tormen & Burstein 1994, in preparation), we will these galaxies to redefine the IRTF relation, re-examine certain kinds of statistical corrections (such as for internal extinction), define groups and clusters within these data in a manner consistent with the rest of the Mark III data set, and re-examine the velocity field within 3000 km s$^{-1}$.

§2 presents the published H magnitude data as observed by the Aaronson et al.

collaboration, together with the optical data we will use for the present analysis. In §3 we detail the methodology employed to define growth curves as a function of galaxy morphological type, including the statistical corrections to optical diameters and the relation of axial ratio to inclination. From this analysis, the full data set is derived and presented. Tests of possible systematics in the derived $H^g_{-0.5}$ magnitudes with other observed properties of spirals are given in §4, together with comparisons with the original Aaronson et al. data set. This analysis is summarized in §5. The main data products of this study are given in Tables 2, 3 and 4. These data will be combined with the other data files relating to the peculiar velocities of nearly 2700 other galaxies to form the *Mark III Catalog of Galaxy Peculiar Velocities* (assembled in collaboration with J. Willick, S. Courteau, S.M. Faber, A. Dressler and A. Dekel). The Mark III Catalog will be given to the National Space Science Data Center (NSSDC) for computer-accessible distribution.

## 2. Published data

### 2.1. H-band aperture magnitude measurements

H magnitude observations of the Aaronson et al. survey are published in eight of their papers from 1979 to 1989, plus the Ph.D. thesis of Marc Aaronson. Galaxies are included in this sample independent of their apparent axial ratios, as we must first define the H mag growth curves, and we do not wish to exclude any kind of galaxy *a priori* from this exercise. Any inclination-dependent effect of the growth curves of these galaxies can be tested *ex post facto* (§4). These nine sources of data are given in Table 1, together with the number of H aperture measurements published and the number code assigned to that paper for our analysis.

We found that of the 1816 H magnitudes published, 85 were duplicated among these papers: 65 between papers A4 and A16, 18 between Aaronson's 1977 Ph.D. thesis and paper A7, and two between papers A7 and A13. In most cases of duplication, the H magnitudes listed were the same; where they were different we took the value quoted in the most recently published paper. This leaves a total of 1731 H magnitude aperture measurements for 665 galaxies, which are listed in Table 2. The columns in Table 2 are as follows: Column **1** (Name) is the name of the galaxy. Column **2** (S) is the source from which the observation was obtained, papers numerically coded as discussed in §1. Column **3** (T) is the photometric numerical type of the galaxy, which defines the growth curve used to fit this galaxy, as discussed in §3. Column **4** (Hmag) is the H-band magnitude measured within aperture A. Column **5** (Err) is the quoted error in the H-band magnitude. Column **6** (log A/D) is the ratio, $\log(A/D_g)$, giving the aperture size in terms of its ratio to the fiducial optical diameter, $D_g$. The values for $\log D_g$ are tabulated in Table 4 (and their derivation discussed in §2.2). Column **7** (Dif) is the difference, in magnitudes, between the individual aperture measurement minus the fitted growth curve for each galaxy as determined in this paper (§3).

In the process of compiling this list we found three errors in the published lists (a very small number for such voluminous tables):

1) There is an atypical typographical error in Table 1 of paper A14. The data listed in column (8) of this table, which gives the aperture size of the observations, apparently was slipped by one row at the entry given for NGC 918. This can be seen by the fact that the larger aperture of the two NGC 918 observations (70.3″) is listed first and corresponds to a fainter magnitude (by 0.34 mag) than the second aperture listed (22.5″). From that point on in this table, the aperture values are offset by one row from the magnitude values, with a blank entry given in this column in the last row. Once this error was found, the correct aperture sizes were assigned to observations for the 25 galaxies listed after NGC 918. After examination of the magnitude difference between the two NGC 918 observations, compared to its optical diameter and other galaxies of similar size, it was decided to assign an aperture size of 54.4″ to the H magnitude observation of NGC 918 for which the aperture size is missing.

2) An apparent error appears in paper A7 for the aperture magnitudes given for NGC 5798. For all galaxies except NGC 5798, the smallest aperture of observation is given first. Only for NGC 5798 is the smallest aperture given second, and the observed H magnitude in the smaller aperture is 0.64 mag *brighter* than that given for the larger aperture. Here we assume that this is a typographical error, and we have reversed the apertures for the given magnitudes for NGC 5798.

3) In paper A13 (Aaronson et al. 1989), the original $H_{-0.5}$ magnitudes quoted in paper A10 (Aaronson et al. 1986) for cluster galaxies were modified when the optical diameter system was modified. Only one galaxy, UGC 12361, has its original H magnitude changed by more than several tenths, from 14.35 (original) to 13.45 (adopted). Since the fiducial diameter for UGC 12361 was made smaller, one would expect the adopted H magnitude to be fainter, not brighter. Moreover, the two H magnitude observations for this galaxy are 14.09 and 14.21 at log $(A/D_g)$ = -0.43, ruling out 13.45 as a possible value of $H_{-0.5}$. We assume that the adopted value of 13.45 listed in paper A13 was entered in error, and the correct value of $H_{-0.5}$ for UGC 12361 from paper A10 should be 14.45.

2.2. Optical Data

Of the 665 galaxies considered here, 639 have values of log $D_{25}$ (the $25^{th}$ mag arcsec$^{-2}$

isophotal diameter) and $\log R_{25} = \log(a/b)$ tabulated in the RC3. Hence, the RC3 data form the basis of our optical data for the present paper. The reader is referred to the introduction of the RC3 for a detailed description of the derivation of these quantities. There are three ways in which the present assumptions concerning working with the optical data differ from those used by Aaronson et al.: (a) transformation of data from different galaxy catalogs to a self-consistent standard system; (b) correction of diameters for internal and Galactic extinction and (c) transformation of observed axial ratios to line-of-sight inclinations (which primarily results in changes to corrected HI 21 cm velocity line widths).

(a) We agree with the prescriptions used by the RC3 for transforming the diameters and axial ratios from the *Uppsala General Catalogue* (Nilson 1973; hereafter UGC), the *ESO/Uppsala Survey of the ESO(B) Atlas* (Lauberts 1982; hereafter ESO) and the *Surface Photometry Catalog of the ESO-Uppsala Galaxies* (Lauberts & Valentijn 1989; hereafter ESO-LV) to a standard RC3 system. The transformations used for these data in the RC3 are significantly different from those used in the *Second Reference Catalog of Bright Galaxies* (de Vaucouleurs, de Vaucouleurs and Corwin 1976; hereafter RC2), which was the basis for the prescriptions used by Aaronson et al. for all of their data. Specifically: The RC2 assumes that $\log D_{25}(RC2) = 0.11 + 0.92 \log D_{UGC}$ and $\log R_{25}(RC2) = 0.894 \log R_{UGC}$. In contrast, the RC3 assumes that $\log D_{25}(RC3) = -0.038 + 1.00 \log D_{UGC}$ and $\log R_{25}(RC3) = 0.98 \log R_{UGC}$. (While no transformations for ESO or ESO-LV data are given in the RC2, Aaronson et al. assumed that similar relationships held for the original ESO diameters and axial ratios; Aaronson et al. 1982b). Almost all of the optical diameters and axial ratios used for the galaxies in the Aaronson et al. survey came either from the RC2 itself, or were transformed to the RC2 system from the UGC or ESO catalogs. Hence, the differences between the RC2 and the RC3 prescriptions directly translate into the same differences between the diameters and axial ratios used here, and those used in the Aaronson et al. papers.

(b) We agree with the prescriptions given in the RC3 for correcting $\log D_{25}$ for inclination-dependent extinction effects. Specifically, we agree that there is no change in $\log D_{25}(RC3)$ with inclination for the UGC, ESO-LV or RC3 data, a result obtained independently by one of us for UGC blue diameters in another paper (Burstein, Haynes and Faber 1991). In contrast, the Aaronson et al. survey modified the RC2 prescription that $\log D_{25}(0)(RC2) = \log D_{25}(RC2) - 0.235 \log R_{25}(RC2)$, limiting the largest possible value of $0.235 \log R_{25}(RC2)$ to 0.15.

We differ with both the RC2 and the RC3 on the correction of optical diameter for Galactic extinction, $A_g$ (using their notation that the subscript g refers to correction for Galactic extinction), preferring the prescription of Rubin et al. (1982):

$\log D_{25}^0 \equiv \log D_g = \log D_{25}(RC3) - \log[1 - (A_g/3.35)]$. (The subscript 'g' is used for this diameter to easily distinguish it from the subscript '1' used by Aaronson et al. for their fiducial optical diameter.) There is little difference between the two extinction-correction methods for galaxies with low galactic extinction, as is the case for the majority of the Aaronson et al. sample. However in specific cases, that of M31 in particular, the difference is significant at the 0.02 dex level. The correction to the observed diameter for redshift is $\leq 0.01$ dex for the galaxies in this sample, a value that is small compared to the errors in the catalog diameters for these galaxies ($\sim 0.1$ dex). Hence no redshift correction is applied to $\log D_g$.

(c) The Aaronson et al. survey used the standard formulation for an oblate spheroid to translate axial ratio (a/b) to predict an initial value of inclination: $\cos^2 i = [(b/a)^2 - q_0^2]/[1.0 - q_0^2]$, where $q_0 = 0.2$ is assumed for the intrinsic flattening of spiral galaxies. In their original paper, Aaronson et al. (Aaronson, Huchra & Mould 1979) correct this inclination value by adding 3°, but then they seem to reject this correction in a later paper (Bothun et al. 1985). In the present paper, we use the same initial formula, but we do *not* add 3°.

To summarize the differences in assumptions used for the present set of optical data and for the optical data used by Aaronson et al. in their survey: (i) Differences in galaxy axial ratios will be correlated with axial ratios in the sense that present data will give galaxies progressively higher axial ratios than in the Aaronson et al. papers. (ii) Diameters of all galaxies will generally be larger in the present sample than in the original Aaronson et al. sample owing to differences in assumed corrections for internal extinction. Differences in tranformations of UGC and ESO-LV diameters will lead to additional increases in diameters for most galaxies. (iii) Inclinations derived from identical axial ratios as given by Aaronson et al. will be, in some cases, smaller by 3° in the present sample.

Of the 26 galaxies in this sample without RC3 data, 15 do not have diameters and axial ratios available from any source except the Aaronson et al. survey, ten have these data available from the ESO and ESO-LV catalogs and one galaxy, Z119-019, has no available diameter from these catalogs or from the Aaronson et al. database. The 15 galaxies with only Aaronson et al. data are all from the cluster survey of Aaronson et al. (1986) (and are noted in Table 4). For these galaxies we assume the Aaronson et al. values and *un*-correct for diameter and extinction according to the differences in the prescriptions detailed above. In the case of the ten galaxies with ESO and ESO-LV data, we apply the same prescriptions as used for the RC3 data.

2.3. HI Velocity Widths

The effect of differences in axial ratios (§2.2) is that the inclination-corrected HI (21 cm) velocity widths quoted for galaxies will not necessarily be the same as quoted in the original Aaronson et al. papers. Unfortunately, many of the Aaronson et al. papers quote rotation velocities already corrected for inclination and redshift effects. In this paper we list the "raw" HI velocity widths, as derived by *un*-correcting the quoted HI velocity widths where appropriate for both inclination (i) and redshift (z) (i.e., by multiplying the quoted HI velocity width by the factor $(\sin i)(1+z)$). We also list the corrected HI velocity widths by dividing by the same factor, but now using the inclinations quoted in this paper. For clarity, the summary data table (Table 4; §3.3) also gives the original Aaronson et al. inclination and quoted (corrected) HI velocity widths for each galaxy.

## 3. Determination of $H^g_{-0.5}$

3.1 Growth Curve Analysis

Growth curves define how the integrated H magnitude of a galaxy increases with the ratio of the aperture size A of an observation to the fiducial diameter (in our case, the isophotal diameter $D_g$). To keep the relationship in terms of logarithms, it is then common to take the log of this ratio, here denoted by $\log(A/D_g)$. In this manner, the integrated luminosity distributions of each galaxy can be compared to other galaxies of similar form, but of different intrinsic size.

The original Aaronson et al. definition of the $H_{-0.5}$ magnitude is based on the data set comprised of the data presented in Aaronson, Huchra & Mould (1979) and in the Ph.D. thesis of Marc Aaronson. Standard H magnitude growth curves are derived from these 289 aperture measurements for spiral galaxies divided into three groups of Hubble types: S0/a,Sa,Sab; Sb,Sbc; Sc,Scd,Sd. Obviously, much more data is now available to define the optically-normalized H-band growth curves of spirals. In addition, many additional galaxies with data now have well-defined Hubble types from the RC3. For these reasons, and others discussed in the Introduction (§1), we proceed with a recalibration of H magnitude growth curves in a manner that is consistent with the data from the full Aaronson et al. survey.

Following Aaronson et al., we assume that a fiducial H magnitude can be defined using H passband aperture measurements and B mag isophotal optical diameters, with optical diameters and axial ratios defined as in §2. Also following Aaronson et al., we derive fiducial $H_{-0.5}$ magnitudes by interpolating (and, if necessary, extrapolating) the H magnitude to the aperture value $\log(A/D_g) = -0.5$, which corresponds to a diameter of 1/3 of the corrected optical isophotal diameter. However, Aaronson et al. defined the final $H_{-0.5}$ magnitudes by direct interpolation of the observed growth curve, and used the standard growth curves only when extrapolation of the data was needed (see paper A2). Differently from them, we always derive the fiducial $H_{-0.5}$ from interpolation or extrapolation of the reference growth curves.

First we need to define our standard growth curves. We divide the sample of 664 galaxies having optical diameters into four classes of photometric types, numbered **1** to **4** (Table 2, coded as "T" there), based on the RC3 Hubble classification notation: (S0, Sa & Sab) = Type **1**; (Sb & Sbc) = Type **2**; (Sc & Scd) = Type **3**; and (Sd to Im) = Type **4**. Types 1 and 2 are the same as the first two subclasses of Aaronson et al. (1979); Types 3 and 4 expand the third subclass of Aaronson et al. into two subclasses, one of which incorporates irregular galaxies. We then proceed to define the reference growth curves for each photometric type in an hierarchical/iterative fashion:

**i.** Galaxies are selected having the most observations at different aperture sizes; each photometric type has at least three galaxies with five or more observations. A preliminary model growth curve is constructed for each photometric type by graphically matching the individual galaxy growth curves and by fitting them together by eye. This initial fit is carried out in the interval $-1.5 \leq \log(A/D_g) \leq -0.3$ for Types 1 and 2, and with an upper limit of –0.26 for Types 3 and 4. All curves are normalized to assume zero value at $\log(A/D_g) = -0.5$, to be consistent with the original Aaronson et al. definition.

**ii.** The observed growth curves of all galaxies with four or more separate observations are least–squares fit to this preliminary curve, as follows. Let $x_1, x_2, \ldots, x_N$ be the normalized apertures observed for one galaxy, and $y_1, y_2, \ldots, y_N$ the corresponding H magnitude values. Let $y = g_k(x)$, $k = 1, \ldots, 4$ denote the model growth curves for the different photometric types, and $w_i$ denote the weight assigned to observation $i$. We define our model as

$$y(x, H_{-0.5}) = g_k(x) + H_{-0.5} \tag{1}$$

and minimize the quantity

$$\sum_{i=1}^{N} [y_i - y(x_i, H_{-0.5})]^2 w_i \tag{2}$$

with respect to the free parameter $H_{-0.5}$. This gives the solution

$$\hat{H}_{-0.5} = \frac{\sum_{i=1}^{N} [y_i - g_k(x_i)] w_i}{\sum_{i=1}^{N} w_i} \tag{3}$$

telling us that the estimated $H_{-0.5}$ magnitude of the galaxy, $\hat{H}_{-0.5}$, is the weighted average distance of the observed growth curve to the model growth curve. The preliminary model growth curve is refined by correcting it by the net values of the residuals $\left(y_i - g_k(x_i) - \hat{H}_{-0.5}\right)$ for the individual galaxy growth curves involved in its definition.

The process of **i.** and **ii.** is iterated several times to remove obvious outlying observations or pathological cases among the galaxy growth curves. The fiducial growth curve is modified by hand in each iteration. (It is our experience that the eye is a better judge of fit when first defining these growth curves, while least-squares tests can later evaluate goodness of fit.)

**iii.** The whole process is repeated again using observed growth curves for galaxies with three or more observations. The growth curves are smoothly extended to $\log(A/D_g) = 0.0$.

Steps **ii** and **iii** are done twice, first with equal weights for all observations: $w_i = 1$, $i = 1, 2, \ldots, N$ then with weights according to quoted errors: $w_i = \sigma_i^{-2}$, $i = 1, 2, \ldots, N$. Negligible differences in predicted values of the H magnitudes are found between these two methods. We adopt the unweighted method for two reasons: First, it makes fewer assumptions about the data and is the easier to reproduce. Second, errors in optical diameters produce comparable effects as errors in H magnitude observations: Errors exist in $\log D_g$ at the level of 0.05 to 0.1 dex, which translate to errors of $\sim 0.1 - 0.2$ mag in $H_{-0.5}$ (§4). The standard H magnitude growth curves defined by this process are tabulated in Table 3, and plotted in Figure 1. 1702 of the original 1731 observations are used to derive the fiducial H magnitudes presented here. 26 observations excluded had values of $\log(A/D_g) < -1.50$ (including 10 observations for NGC 224 alone); one observation had an error of 0.18 mag (for E154-23); and the two observations for Z119019 were not used as this galaxy does not have a fiducial diameter.

**iv.** Residuals from the growth curve fits are plotted for each galaxy (similar to those given in Burstein et al. (1987); their Figure 3) as a function of $\log(A/D_g)$ at several stages in the process. These residuals typically show one of three kinds of patterns: i) a good fit, indicating the standard growth curve is a satisfactory match; ii) intrinsic scatter among the observations which no standard growth curve can fit; and iii) a systematic trend as a function of $\log(A/D_g)$. If the systematic trend indicates that a different photometric type would fit the growth curve better, another photometric type is used. Such is the case for 12 galaxies (noted in Table 4). Figure 2 plots the residuals individually for 120 galaxies with four or more observations as illustrations of this process. As with similar plots in Burstein et al. (1987), the number code for the paper source of the H magnitude observation is used as the plotting symbol. Figure 3 plots the residuals of the growth curve fits for galaxies with three or more observations as a function of $\log(A/D_g)$ for each type of growth curve. (Three as the minimum number of observations per galaxy for Figure 3 and later figures is based on the philosophy that only 1 or 2 observations per galaxy do not help to define systematic trends in growth curve fitting. The growth curve residuals are listed for all observations used for all galaxies in Column 7 of Table 2, under the heading "Dif.")

One result of the above methodology is that the standard growth curves finally adopted do not have to perfectly fit the growth curves of the galaxies that defined the preliminary version. Indeed, systematic differences of $\sim 0.1 - 0.2$ mag are seen between the growth curves of many of the galaxies that initially defined the Type 2 and Type 3 growth curves and the growth curves that were finally adopted for these types. Most of the deviations that are seen occur for $\log(A/D_g) < -1.0$, while the majority of the galaxies in this sample have growth curves defined primarily for $\log(A/D_g) > -1.0$, as can be seen by inspection of Table 2 and Figure 3.

### 3.2 $H^g_{-0.5}$: Quality and Selection

The values of $H_{-0.5}$ that result from the growth curve fitting are corrected for Galactic extinction by $-0.1A_g$ (separate from the extinction correction to diameter, already applied in the definition of $D_g$; §2.2). In almost all cases, this correction factor is smaller than 0.1 mag. The resulting values we term $H^g_{-0.5}$, as we have not yet attempted to correct these data for the direct dependence on inclination of the derived H magnitude. Until that correction is made, a redshift-dependent K-correction for H magnitude will also not be made. It is our judgement that an axial ratio-dependent correction to $H^g_{-0.5}$ is best determined from the scatter in the IRTF relation itself. While internal extinction effects at H magnitude are expected to be small, per se, this test should also discover if our assumption about diameter dependence on axial ratio is correct. It will also be sensitive to inclination-dependent issues related to calculating HI velocity widths. The re-derivation of the IRTF relation will be done in the second paper related to these data (Willick, Tormen and Burstein 1994).

The quality of fit to the standard growth curve for each galaxy was assessed through examination of the residuals of the fit plotted versus $\log(A/D_g)$, of which the plots in Fig. 2 are examples. In this dataset 65% of the H magnitudes have a quoted error of 0.03 mag and only 1% have errors >0.1 mag (Table 2). It is our experience that the data with larger quoted errors do, indeed, show more scatter in the residual plots. In addition, it appears that the H magnitude data suffer, as do optical photometry (cf. Burstein et al. 1987) from "wild points", observations that obviously disagree with other self-consistent observations for reason or reasons unknown. Such wild points can be a product of many things that can happen at the telescope (point at the wrong galaxy; mis-center galaxy in aperture; galaxy not in focus; dome partially occults telescope), and seem to exist in any sample of galaxy photometry taken point-by-point. Fortunately, there are few such wild points in the Aaronson et al. sample.

The quality assessments made here are based primarily on the residual plots for each galaxy. We also quadratically add the ~0.15 mag error resulting from errors in $D_g$ (§3.1; see also below) into the error estimate for the growth curve fits. The qualities are given on a 1-2-3 scale, analogous to what was done for the 7 Samurai data set (Burstein et al. 1987; Faber et al. 1989). Quality 1 indicates an error of <0.20 mag for $H^g_{-0.5}$, and is given when the growth curve fit had all aperture magnitudes (at least two) in agreement with the reference growth curve to better than 0.1 mag near the value $\log(A/D_g) = -0.5$. Quality 2 indicates an error of 0.20–0.35 mag and is given when there is a disagreement among two or more aperture measurements at a level of 0.1 to 0.2 mag and an extrapolation $\geq 0.2$ dex in $\log(A/D_g)$ is required for an otherwise well-defined growth curve, or there are other indications the value of $D_g$ might be suspect. Quality 3 indicates an error of >0.35 mag and results if: **a)** there is a disagreement >0.2 mag among two or more aperture measurements; **b)** there is only one aperture measurement within the interval of definition of the reference growth curve; or **c)** the quoted value of $\log D_g$ is known to be in error by more than 0.1 dex (as is the case for some of the cluster galaxies without RC3 diameters).

In order to be able to generate a final list of galaxies for IRTF studies one should, as did Aaronson et al., pare the full list down based on a series of objective criteria. Two of these exclusion criteria we share in common with Aaronson et al. Galaxies are excluded from IRTF studies if they have either i) inclinations $\leq 45°$ to the line-of-sight (in practical terms, $\log(a/b) < 0.15$) or ii) no quoted values or poorly-determined values of HI velocity widths given by Aaronson et al. In addition, we also exclude galaxies classified as "peculiar" (P), I0 or lenticular (L), using the RC3 or ESO Hubble types.

Based on these criteria, we exclude 39 galaxies from the sample of 363 "field" galaxies (some of these galaxies are actually in groups and clusters) that were distributed in the *Mark II Catalog of Galaxy Peculiar Velocities*. 14 galaxies have no published HI velocity widths, 9 galaxies are too face-on with present values of axial ratios and 16 are otherwise typed P (peculiar), I0 (Irr II) or L (S0). (It should be noted that for Aaronson et al. field galaxies, the *Mark II Catalog* included data that were privately communicated to DB by Marc Aaronson in 1983. Apparently the data for 14 galaxies in that 1983 list were later eliminated by the Aaronson et al. survey from IRTF studies. We also eliminate these galaxies.) Of the 204 "cluster" galaxies in the *Mark II Catalog*, 12 were excluded that are typed peculiar, Irr II or S0 and 6 were excluded that are too face-on. Ten galaxies are *added* to the field sample that were excluded by Aaronson et al. for reasons pertaining to non-cluster membership, but the optical data for which is otherwise satisfactory and Aaronson et al. published reliable values of HI velocity widths. Two galaxies are added to the original cluster list. In all, 522 galaxies are included in the sample for further IRTF studies and 143 galaxies are rejected: 101 that have no quoted HI velocity widths in the Aaronson et al. papers; 15 galaxies too face-on and 28 galaxies otherwise classified as peculiar, Irr II or S0. A code is given in Table 4 as to whether a galaxy is included (G) or excluded (N, F or P) from the final sample.

3.3 Data Summary

Table 4 summarizes the assumed and derived data for each galaxy in this sample, by column: Column **1** (Name) is the standard name of the galaxy, in order of preference for NGC, IC, UGC, ESO or Zwicky designations. This is the same name as used in Table 2. Column **2** (PGC#) is the *Principal Galaxy Catalog* number (Paturel et al. 1989), taken either from the RC3 or, in the cases for 13 galaxies not in the RC3, directly from the PGC. Galaxies in both Table 2 and Table 4 are listed in order of the PGC number. Columns **3** (R.A.) and **4** (Dec) are the 1950 values of Right Ascension (hr, min, sec with no spaces inbetween) and Declination (degree, min, sec with no spaces inbetween) for each galaxy, taken either from the RC3, ESO or, in the case of four galaxies, from the PGC. Column **5** (C) is a numerical code for the cluster or group to which this galaxy was assigned by Aaronson et al., as explained in the Notes to the Table 4. Galaxies with "cluster" codes between 1 and 20 are in the original "cluster" Aaronson et al. sample as defined in the *Mark II Catalog of Galaxy Peculiar Velocities*. Galaxies with cluster codes of -1 were not in the *Mark II Catalog* and did not have their group membership defined by Aaronson et al. Galaxies with cluster codes of 0 were in the original *Mark II Catalog* but were not placed into a cluster or group by Aaronson et al. Galaxies with cluster codes of 50 to 80 are clusters and groups originally contained in the "field" sample of the *Mark II catalog*.

Column **6** (RC3t) is the Hubble type taken from either the RC3 or ESO catalogs or, in the case of Z097058, Z476112 and Z119019, from Aaronson et al. This format is defined in the RC3. Column **7** (T) is the photometric type assigned to each galaxy for growth curve analysis (§3.1 and Table 2). The 12 galaxies for which the photometric types were changed are noted with an asterisk. Column **8** (EXT = $A_g$) is the blue magnitude Galactic extinction, defined as $4.0 \times E(B - V)$, as defined by Burstein and Heiles (1978). Column **9** (log Dg) is $\log D_g$ (in standard RC3 units of $0.1'$), the fiducial corrected isophotal diameter used in this analysis. Most diameters come from the RC3; several come from the ESO catalog and 15 were derived from quoted Aaronson et al. values. These latter 15 galaxies have their values of $\log D_g$ noted with an asterisk. Column **10** (log R) is $\log R_{25} = \log(a/b)$, taken mostly from the RC3. The 15 axial ratios not obtained from either the RC3 or ESO catalogs are marked with an asterisk. Column **11** (i) is the inclination (in degrees) adopted for the given axial ratio in Column **10**.

Column **12** (Vhel) is the heliocentric radial velocity (km s$^{-1}$) either quoted by the Aaronson et al. survey or obtained from the NASA/IPAC Extragalactic Database (for 23 galaxies). Column **13** (dVr) is the "raw" HI velocity width *not* corrected for inclination. Column **14** (log dV) is the logarithm of the inclination-corrected HI velocity width, using the inclination given in Column **11** and corrected for heliocentric redshift, as discussed in §2.3. Column **15** (H_g) is the fiducial H magnitude defined by the present analysis, corrected only for Galactic extinction, $H^g_{-0.5}$. Column **16** (Q) is the quality parameter for $H^g_{-0.5}$, as discussed in §3.2. Column **17** (N) is the number of H magnitude observations that defined the growth curve from which $H^g_{-0.5}$ was derived. Column **18** (U) has the code for inclusion or exclusion of a galaxy in the final IRTF sample: G = included in the final IRTF sample; F = too face-on; N = no HI velocity width profile in Aaronson et al. papers; P = Hubble type of P, I0 or L. Columns **19** through **22** contain the original Aaronson et al. data where available. Column **19** (I2) is the inclination (in degrees) taken from the Aaronson et al. survey. Column **20** (AHMdV) is the logarithm of HI velocity width corrected for the inclination given in Column **19** and heliocentric redshift. This value can be compared to that from the present survey, given in Column **14**. Column **21** (log D1) is the fiducial corrected optical diameter defined by Aaronson et al, $\log D_1$, in standard RC3 units of $0.1'$. Finally, Column **22** (H_0) is the corrected H magnitude used by Aaronson et al., $H^0_{-0.5}$.

For those who wish to use this paper for more detailed reference, we give here the 11 galaxies which have two different names as used in the original Aaronson et al. papers (excluding MCG designations): NGC 6690 = NGC 6689 = UGC 11300; NGC 7361 = IC 5237; U 8861 = Z074010; U 8951 = Z074060; Z097180 = Z098002; UGC 6525 = Z126083; Z406031 = A2312+07; UGC 12855 = Z477033; U 501 = Z501021; and IC 2308 = Z089006.

## 4. Statistical Tests

### 4.1. Internal Tests

Both $\log D_g$ and $\log(a/b)$ are important parameters in determining the value of $H^g_{-0.5}$ derived from the growth curve fits. The effect of errors in diameters is direct, as such errors translate directly into systematic errors in $\log(A/D_g)$ for galaxies so affected. Similarly, any errors in the correction of diameters for Galactic extinction would have the same effect. In the case of axial ratios, the effect is indirect, as axial ratios go into determining several parameters that determine the derived value of $H^g_{-0.5}$ as well as corrected HI velocity width, including: inclination, internal extinction correction for diameter, and internal extinction correction for magnitude. (Note that of the three corrections correlated with axial ratio, only the first two have been applied to these data. Internal extinction corrections will be determined when the Tully-Fisher relationship for this sample is determined, §3.2.)

To test for each of these effects, we have binned the sample into three intervals in diameter, axial ratio and Galactic extinction and into the four photometric types. For the galaxies that fall into each interval/type, we plot the aperture magnitude residuals, termed here "growth curve residuals," of their growth curve fits (Table 2) as functions of $\log(A/D_g)$ in Figures 4 through 7. Galaxies of all quality parameters are plotted, but galaxies with less than three observations are excluded, as with Figure 3.

Inspection of the plots in Figures 4–6 show some indication of systematic growth curve residuals of amplitude $\sim 0.05$ mag (e.g., edge-on galaxies of Types 1 and 2 in Figure 5, or Type 1 galaxies of intermediate diameters in Figure 4). The strengths of these systematic trends are about at the level (0.05 mag) of accuracy at which the standard growth curves are probably determined. Hence, no attempt was made to "fine-tune" the growth curves for these other effects, as it was felt that such fine-tuning is beyond the capability of the present data set.

To complete these internal comparisons, in Figure 7 we plot the growth curve residuals for galaxies independent of Type, but grouped by quality of fit. By inspection it is obvious that: i) there is no systematic trend in the residuals when all of the galaxies are grouped together and ii) our quality parameter is correlated with the amount of scatter in the growth curve fit. In sum, these internal tests indicate that the optically-normalized H magnitude growth curves determined here (Table 3) are accurate to 0.05 mag over the range $-1.5 \leq \log(A/D_g) \leq -0.2$.

4.2. Comparison to the Original Aaronson et al. Data

As discussed in §2.2, our diameter system differs systematically from that of Aaronson et al., primarily as a function of axial ratio. This can be seen explicitly in Figure 8, which plots the logarithmic difference in optical diameters, $\log D_g - \log D_1$ (our values minus those of Aaronson et al.) versus the present value for $\log(a/b)$ separately for the Aaronson et al. field survey (left hand side of the figure) and for the cluster survey (right hand side), defined in the *Mark II Catalog* sense, as a function of photometric type (§3.3). Data for all galaxies in the sample are plotted here (not just those with three or more observations as in previous figures) if a value for $\log D_1$ exists (i.e., not all galaxies in the field survey have published values of $\log D_1$). The line drawn in each plot of Figure 8 is the predicted correlation that would exist if the galaxies perfectly followed the two different axial ratio corrections used by our analysis and that of Aaronson et al. — $\log(D_g/D_1) = 0.235 \log(a/b)$ for $\log(a/b) < 0.63$, and $\log(D_g/D_1) = 0.15$ for $\log(a/b) \geq 0.63$ (§2.2).

As is evident, most of the trend that exists in both field and cluster sample is adequately explained by the difference in assumed dependence of angular diameter on internal extinction. The ±0.1 dex scatter in Figure 8, relative to the predicted relationship, is consistent with known random errors in the transformations of UGC and ESO data to a standard system (§2.2) in both the present data set and that of Aaronson et al.

Differences in fiducial diameters will naturally lead to differences in fiducial magnitudes derived from growth curve fitting. Thus, given our diameters are larger on average, it is no surprise that $H^g_{-0.5}$ values are brighter than $H^0_{-0.5}$ values quoted by Aaronson et al. by $0.21 \pm 0.007$ mag. Figure 9 shows explicitly that this difference is due to the different diameter systems by plotting the $\Delta H = H^g_{-0.5} - H^0_{-0.5}$ versus $\log D_g - \log D_1$. In analogy to Figure 8, the data in Figure 9 are divided both by photometric type and into field and cluster samples.

It is evident from inspection that: i) The differences in H magnitude are well correlated with axial ratio for the field galaxies, but somewhat less well-correlated for cluster galaxies. ii) Galaxies of photometric types 2 and 4 have more scatter than for Types 1 and 3. The larger scatter for cluster data is not surprising as these are among the fainter and apparently smaller galaxies in this sample. The fact that the photometric type 4 was not used by Aaronson et al. leads to the increased scatter for galaxies assigned this type in the present sample.

## 5. Summary

The H magnitude aperture measurements published by the Aaronson et al. collaboration in 16 papers over a 10-year period form the cornerstone of our current understanding of the large-scale peculiar velocity field out to distances of $\sim 3000$ km s$^{-1}$. These data have been compiled into one list containing 1731 separate observations for 665 spiral, irregular, S0 and I0 galaxies in the field, small groups and clusters. In the process of compilation duplicate observations have been eliminated and two important typographical errors found in the original papers are corrected.

The full data set permit the sample to be divided into four photometric types (S0,Sa,Sab; Sb,Sbc; Sc,Scd; Sd,Sdm,Im), as opposed to the original three types used by Aaronson et al. (1979). Moreover, the RC3 gives optical isophotal diameters for 96% of these galaxies, making it possible to derive internally-consistent growth curves for both field and cluster galaxies. Both the isophotal diameter system and axial ratio system defined by the RC3 and this paper ($\log D_g$) differ systematically from those used by Aaronson et al. ($\log D_1$), which were based on the RC2 precepts.

These differences are quantitatively addressed in the present analysis, both as to their sources and as to the effects on the eventual derivation of fiducial H magnitudes.

Tests have been made to look for systematic effects as a function of galaxy diameter, axial ratio, Galactic extinction or photometric type. All tests indicate that the H magnitude standard growth curves have been determined to an accuracy of 0.05 mag for $-1.5 \leq \log(A/D_g) \leq -0.2$.

The main results of this analysis are presented in terms of three tables: Table 2 gives the 1731 H magnitude observations, together with the values of $\log(A/D_g)$ and the growth curve residual for each aperture measurement. Table 3 gives the growth curves used for this analysis, which should be useful for comparison to H magnitude growth curves derived using H magnitude diameters (as opposed to optical diameters). Table 4 gives a summary of the data derived for this analysis, as well as some of the analogous data given by Aaronson et al.

The present values of fiducial H magnitude are given the superscript 'g' to denote the fact that these magnitudes are only corrected for Galactic extinction, and are not corrected for either redshift or possible inclination dependent effects. It is our determination that these latter two corrections be derived from a detailed reanalysis of the near infrared Tully-Fisher relationship defined by these new data. This latter analysis will be done in our second paper (Willick, Tormen & Burstein 1994), in which we also use these data to reexamine the large-scale, local peculiar velocity field.

This reanalysis of the Aaronson et al. H magnitude data set has been done as part of the more general assemblage of internally-consistent peculiar velocities of galaxies that will go into making the *Mark III Catalog of Galaxy Peculiar Velocities*. Altogether, the *Mark III Catalog* will contain distances for 1000 galaxies from the *Mark II Catalog* together with distances for over 2300 new galaxies, all reanalyzed according to the precepts of Willick (1991). The *Mark III Catalogue* is a collaborative effort among the present authors and J.A. Willick, S. Courteau, S.M. Faber, A. Dressler and A. Dekel, and will be distributed to the community through the facilities of the National Space Science Data Center (NSSDC).

**Acknowledgements**

We would like to thank Marc Aaronson, Jeremy Mould and John Huchra for assistance in helping us assemble this database, and the referee for useful editorial comments. Use of the NASA/IPAC Extragalactic Database (NED) is gratefully acknowledged. NED is operated by IPAC under a contract from NASA. This research was supported by NSF Grant AST 90-16930 to DB and by the Italian MURST to GT.

Table 1

Sources of H Magnitude Aperture Photometry

| Aaronson et al. Paper | Paper # | # Obs | Duplicates |
|---|---|---|---|
| Aaronson, Huchra, Mould 1979 | 1 | 81 | |
| Aaronson, Mould, Huchra 1980 | 2 | 55 | |
| Mould, Aaronson, Huchra 1980 | 3 | 28 | |
| Aaronson et al. 1980 | 4 | 93 | Paper 16: 65 obs |
| Aaronson et al. 1981 | 5 | 35 | |
| Aaronson et al. 1982b | 7 | 634 | Paper 13: 2 obs; Paper 99: 18 obs |
| Aaronson et al. 1989 | 13 | 216 | |
| Bothun et al. 1984 | 14 | 48 | |
| Bothun et al. 1985 | 16 | 418 | |
| Aaronson Ph. D. Thesis | 99 | 208 | |

Table 3

Standard H Magnitude Growth Curves

| $\log(A/D_g)$ | Type 1 (S0,Sa,Sb) | Type 2 (Sb,Sbc) | Type 3 (Sc,Sd) | Type 4 (Sdm-Im) |
|---|---|---|---|---|
| -1.500 | 1.918 | 2.052 | 2.495 | 3.155 |
| -1.400 | 1.708 | 1.832 | 2.231 | 2.829 |
| -1.300 | 1.504 | 1.621 | 1.967 | 2.503 |
| -1.200 | 1.304 | 1.411 | 1.703 | 2.177 |
| -1.100 | 1.104 | 1.201 | 1.439 | 1.851 |
| -1.000 | 0.904 | 0.991 | 1.175 | 1.525 |
| -0.950 | 0.809 | 0.891 | 1.043 | 1.362 |
| -0.900 | 0.714 | 0.791 | 0.911 | 1.199 |
| -0.850 | 0.619 | 0.691 | 0.779 | 1.036 |
| -0.800 | 0.524 | 0.591 | 0.647 | 0.878 |
| -0.750 | 0.429 | 0.491 | 0.522 | 0.721 |
| -0.700 | 0.338 | 0.391 | 0.408 | 0.563 |
| -0.650 | 0.248 | 0.291 | 0.301 | 0.412 |
| -0.600 | 0.158 | 0.191 | 0.196 | 0.268 |
| -0.550 | 0.075 | 0.093 | 0.094 | 0.130 |
| -0.500 | 0.000 | 0.000 | 0.000 | 0.000 |
| -0.450 | -0.067 | -0.087 | -0.089 | -0.121 |
| -0.400 | -0.129 | -0.165 | -0.172 | -0.229 |
| -0.350 | -0.189 | -0.235 | -0.249 | -0.322 |
| -0.300 | -0.249 | -0.297 | -0.317 | -0.408 |
| -0.250 | -0.305 | -0.357 | -0.378 | -0.485 |
| -0.200 | -0.360 | -0.416 | -0.433 | -0.554 |
| -0.150 | -0.411 | -0.461 | -0.486 | -0.619 |
| -0.100 | -0.461 | -0.514 | -0.536 | -0.679 |
| -0.050 | -0.510 | -0.564 | -0.582 | -0.736 |
| 0.000 | -0.555 | -0.610 | -0.627 | -0.789 |

# References

Aaronson, M. 1977, Ph.D. thesis, Harvard Univ (A99)

Aaronson, M. 1986, in Galaxy Distances and Deviations from Universal Expansion, eds. B. F. Madore & R. B. Tully (Dordrecht: Reidel), 55 (A9)

Aaronson, M., Bothun, G. D., Mould, J., Huchra, J., Schommer, R. & Cornell, M. 1986, ApJ 302, 536 (A10)

Aaronson, M., Dawe, J., Dickens, R. J., Mould, J., Murray, J. 1981, MNRAS 195, 1P (A5)

Aaronson, M., Huchra, J. & Mould, J. 1979, ApJ 229, 1 (A1)

Aaronson, M., Mould, J., & Huchra, J. 1980, ApJ 237, 655 (A2)

Aaronson, M., Huchra, J., Mould, J., Schechter, P. L. & Tully, R.B. 1982a, ApJ 258, 64 (A6)

Aaronson, M. & Mould, J. 1983, ApJ 265, 1 (A8)

Aaronson, M. & Mould, J. 1986, ApJ 303, 1 (A11)

Aaronson, M., Mould J., Sullivan, W. T. III, Schommer, R. A., Bothun, G. D. 1980, ApJ 239, 12 (A4)

Aaronson, M., et al. 1982b, ApJS 50, 241 (A7)

Aaronson, M., et al. 1989, ApJ 338, 654 (A13)

Bothun, G. D., Aaronson, M., Schommer, R., Huchra, J. & Mould, J. 1984, ApJ 278, 475 (A14)

Bothun, G. D., Aaronson, M., Schommer, R., Mould, J.R., Huchra, J. & Sullivan, W. T. III 1985, ApJS 57, 423 (A16)

Bothun, G. D. & Mould, J. R. 1987, ApJ 313, 629 (A15)

Burstein, D 1990, Rep Prog Phys 53, 421

Burstein, D., Faber, S.M. & Dressler, A. 1990, ApJ 354, 18

Burstein, D., Haynes, M.P. & Faber, S.M. 1991, Nature 353, 515

Burstein, D. & Heiles, C. 1978, ApJ 225, 40

Burstein, D., Davies, R.L., Dressler, A., Faber, S.M., Stone, R.P.S., Lynden-Bell, D., Terlevich, R. & Wegner, G. 1987, ApJS 64, 601

Cornell, M.E., et al. 1987, ApJS 64, 507 (A12)

Faber, S. M. & Burstein, D. 1988, in Large-Scale Motions in the Universe, eds. V. C. Rubin & G. Coyne (Princeon: Princeton Univ Press), 115

Faber, S.M, Wegner, G., Burstein, D., Davies, R.L., Dressler, A., Lynden-Bell, D. & Terlevich, R. et al. 1989, ApJS 69, 763

Lauberts, A. 1982, The ESO/Uppsala Survey of the ESO(B) Atlas, (Garching-bei-München: European Southern Obs) (ESO)

Lauberts, A. & Valentijn, E. A. 1989, The Surface Photometry Catalog of the ESO-Uppsala Galaxies, (Garching-bei-München: European Southern Obs) (ESO-LV)

Mould, J., Aaronson, M., & Huchra, J. 1980, ApJ 238, 458 (A3)

Nilson, P. 1973, Uppsala General Catalog of Galaxies, (Uppsala: Roy Soc Sci Uppsala) (UGC)

Paturel, G., Fouqué, P., Bottinelli, L. & Goughenheim, L. 1989, Catalogue of Principal Galaxies, (Lyon: Obs de Lyon) (PGC)

Rubin, V.C., Ford, W.K. Jr, Thonnard, N. & Burstein, D. 1982, ApJ 261, 439

Shaya, E.J., Tully, R.B & Pierce, M.J 1992, ApJ 391, 16

Tully, R. B. & Fisher, J. R. 1977, A & A 54, 661

de Vaucouleurs, G., de Vaucouleurs, A., Corwin, H.G., Jr., Buta, R. J., Paturel, G. & Fouqué, 1991, The Third Reference Catalog of Bright Galaxies, Vol 1-3, (New York: Springer-Verlag) (RC3)

de Vaucouleurs, G., de Vaucouleurs, A. & Corwin, H.G., Jr. 1976, The Second Reference Catalog of Bright Galaxies, (Austin: Univ of Texas Press) (RC2)

Willick, J. A., Tormen, G. & Burstein, D. 1994, in preparation

# Figure Captions

**Figure 1.** Reference growth curves for the 4 photometric types. In abscissa the apertures are normalized to the isophotal diameter $D_g$; in ordinate $g_k(y)$ ($k = 1, \ldots, 4$) is the predicted magnitude difference to the value measured at $\log(A/D_g) = -0.5$. Note that late morphological types correspond to steeper growth curves for aperture sizes less than $\log(A/D_g) = -0.5$.

**Figure 2.** Residuals from the growth curve fits for 120 galaxies which have four or more observations used for these fits, divided into four sets of 30 galaxies (a,b,c,d). The plotting symbol used is the code number for the source of aperture photometry (see reference list and text). The vertical scale is $\pm 0.35$ mag for each individual plot, and the horizontal scale is $-1.5 \leq \log(A/D_g) \leq 0.0$. For each galaxy is given the PGC number, galaxy name, numerical photometric type (1 to 4), quality parameter for the fit (1 to 3) and the number of aperture measurements used (4 to 9).

**Figure 3.** Residuals of growth curve fits (GC residuals) given in magnitudes, as a function of $\log A/D_g$ for photometric type 1 (a), Type 2 (b), Type 3 (c) and Type 4 (d) galaxies. Only data for galaxies with growth curves defined by three or more observations are plotted. Closed circles denote data for galaxies with Quality 1 $H^g_{-0.5}$ magnitudes; open squares denote Quality 2 data; open circles Quality 3 data.

**Figure 4.** Growth curve residuals (in magnitudes) are plotted versus $\log(A/D_g)$ for the same sample of galaxies as in Figure 3. The twelve panels divide the sample into the four photometric types (each row) and into three intervals of apparent corrected isophotal diameter: The first column is for galaxies with $\log D_g < 1.35$; the second column is for galaxies with $1.35 \leq \log D_g < 1.75$; and the third column is for galaxies with $\log D_g > 1.75$. Plotting symbols denote the Quality parameters of the data, as in Figure 3.

**Figure 5.** The same sample of galaxies as in Figure 3, but here divided by photometric type and axial ratios: The first column is for galaxies with $\log(a/b) < 0.3$; the second column is for galaxies with $0.3 \leq \log(a/b) < 0.6$; and the third column is for galaxies with $0.6 \leq \log(a/b)$. Plotting symbols the same as in Figure 3.

**Figure 6.** The same sample of galaxies as in Figure 3, but here divided by photometric type and Galactic extinction: The first column is for galaxies with $A_g < 0.1$; the second column is for galaxies with $0.1 \leq A_g < 0.3$; and the third column is for galaxies with $0.3 \leq A_g$. Plotting symbols the same as in Figure 3.

**Figure 7.** The same sample of galaxies as in Figure 3, but here divided into the three classes of Quality parameters: (a) Quality 1; (b) Quality 2; (c) Quality 3. Plotting symbols

represent the photometric Types as given in the legend.

**Figure 8.** Corrected isophotal optical diameter used in the present analysis, $\log D_g$ (TB) ratioed to the corrected isophotal diameter used by Aaronson et al., $\log D_1$ (AHM) (TB = Tormen and Burstein; AHM = Aaronson et al.), plotted versus the axial ratio, $\log(a/b)$ used in the present analysis. The data are divided by photometric type and whether they were in the original Aaronson et al. "field" sample or the "cluster" sample (see text). The line drawn in each figure is *not* fitted to these data, but is the predicted relationship between axial ratio and this diameter ratio, based on the different manner in which axial ratio corrections to isophotal diameters were assumed by the present analysis and that of Aaronson et al.

**Figure 9.** Differences in fiducial H magnitude from the present analysis ($H^g_{-0.5}$) and that of Aaronson et al. ($H^0_{-0.5}$), plotted versus $\log(D_g/D_1)$ (TB/AHM). The format is the same as in Figure 8. It is apparent that the differences in H magnitude are well correlated with this diameter ratio for all photometric types and for both field and cluster galaxies. Somewhat weaker correlations exist for Type 2 and Type 4 galaxies and for cluster galaxies relative to field galaxies.